\documentclass[twocolumn,preprintnumbers,amsmath,amsfonts,amssymb,floatfix,aps,pra,superscriptaddress]{revtex4}

\usepackage[utf8]{inputenc}

\usepackage{graphicx}
\usepackage{amssymb}
\usepackage{amsmath}
\usepackage{amsfonts}
\usepackage{bm}
\usepackage{color}
\usepackage{hyperref}

\usepackage{braket}

\newcommand{\be}{\begin{equation}}
\newcommand{\ee}{\end{equation}}

\begin{document}

\title{Downfolding a quantum many-body system: the quasi-1D Fermi polaron}

\author{Lovro Anto Barisic}%
\email{Corresponding author: lovro.barisic@phys.ens.fr}
\affiliation{Laboratoire de physique de l'Ecole Normale sup\'erieure, ENS, Universit\'e PSL, CNRS, Sorbonne Universit\'e, Universit\'e Paris Cit\'e, F-75005 Paris, France.
}%

\author{Giuliano Orso}
\affiliation{Universit\'e Paris Cit\'e, Laboratoire Mat\'eriaux et Ph\'enom\`enes Quantiques (MPQ), CNRS, F-75013, Paris, France}

\author{Kris Van Houcke}%
\affiliation{Laboratoire de physique de l'Ecole Normale sup\'erieure, ENS, Universit\'e PSL, CNRS, Sorbonne Universit\'e, Universit\'e Paris Cit\'e, F-75005 Paris, France.
}%

\author{Fr\'ed\'eric Chevy}%
\affiliation{Laboratoire de physique de l'Ecole Normale sup\'erieure, ENS, Universit\'e PSL, CNRS, Sorbonne Universit\'e, Universit\'e Paris Cit\'e, F-75005 Paris, France.
}%
\affiliation{Institut Universitaire de France (IUF), 75005 Paris, France}

\begin{abstract}

We investigate the properties of an impurity immersed in an ensemble of spin-polarized fermions confined in a tight quantum wire. We use a non-perturbative variational approach that accounts for virtual transverse excitations and regularizes the zero-range interaction. We compute the polaron's energy, effective mass, and spectral weight, and benchmark our results against the exactly solvable Yang-Gaudin model of a purely 1D system. While the two models agree in the weakly interacting regime, we find significant deviations in the strongly attractive limit, including a divergence of the effective mass and evidence for a polaron-to-molecule transition inherited from the underlying 3D physics, which is absent in the purely 1D description. Our work quantifies the accuracy of quasi-1D systems as analog quantum simulators and highlights the emergence of beyond-1D physics even in nominally 1D settings.
\end{abstract}

\maketitle

Quantum simulation is one of the pillars of the quantum technology program \cite{bloch2008many,bloch2012quantum,Altman2021QuantumOpportunities} and its objective is to develop experimental systems capable of reproducing complex mathematical models currently beyond the reach of traditional analytical or numerical approaches. By definition, these systems operate in a regime where comparison with theory is impossible and their certification and the quantification of the accuracy at which they reproduce the targeted equations is a major challenge for the development of next-generation quantum simulators \cite{Hangleiter2017DirectSimulations,Xiao2022IntelligentLearning}. 

In the past decade, experimental platforms based on ultracold Fermi gases near Feshbach resonances have enabled a dramatic leap in our understanding of strongly correlated quantum matter \cite{bloch2008many,bloch2012quantum}, and these groundbreaking experiments have paved the way to the field of precision many-body physics where quantum many-body physics can be tested at an unprecedented level. In particular, the control of the potential landscape in which atoms evolve using optical potentials made it possible to implement experimentally model systems such as the Hubbard model \cite{esslinger2010fermi,gross2017quantum} or to study the impact of low dimensionality on many-body physics \cite{Paredes2004TonksGirardeauLattice,liao2009spin,Orso2007Attractive,parish2007quasi}. These models correspond usually to effective theories describing the low-energy physics of the experimental system, when the typical energies are much smaller than some band gap isolating the low- and high-energy sectors. In the strongly correlated regime, virtual transitions towards higher bands can alter the elimination of the excited degrees of freedom and the downfolding procedure gives rise to novel terms in the Hamiltonian, like density hopping terms in the Hubbard model \cite{Werner2005Interaction-inducedLattices,Jiang2023Density-matrix-renormalization-group-basedHopping} or emergent few-body interactions in lower dimensional systems \cite{Mazets08breakdown,tan10relaxation,pricoupenko19three,  Chevy2022AchievingFermions,Shi2023DimensionalTraps}. 
This mechanism might, for instance, be responsible for the  intriguing results reported in  \cite{Koschorreck2012AttractiveDimensions,Sobirey2022ObservingSuperfluids}, and may play a strong role in the description of the properties of high-Tc superconductors \cite{Jiang2023Density-matrix-renormalization-group-basedHopping}.

In this letter, we explore the case of the so-called quasi-1D Fermi polaron problem that describes an impurity immersed in a 1D Fermi sea of spin-polarized fermions. Impurity problems have been a very popular field of research in ultracold atomic physics over the past decade \cite{Chevy2010Unitary,Koschorreck2012AttractiveDimensions,massignan2014polarons,Parish2023FermiBeyond,Massignan2025PolaronsSemiconductors} and significant theoretical and experimental \cite{nascimbene2009pol,schirotzek2009ofp,Schmidt2012FermiDimensions,Wenz2013FromTime,Ness2020ObservationGas,Yan2019BoilingLiquid,Adlong2020QuasiparticlePolaron,kohstall2011metastability,Vivanco2025ThePolaron} progress have been realized in our understanding of the properties of these systems. In particular, powerful analytical techniques \cite{chevy2006upa,combescot2007nsh,Giraud2009HighlyCase,Moser2017StabilityInteractions,Liu2019VariationalTemperature,Cui2020FermiCoexistence,Peng2021NaturePolarons} have been developed. These approaches are usually based on an expansion over the number of particle-hole excitations and were benchmarked to exact Monte-Carlo simulations \cite{prokofev2008bdm,van2020high}. Here, we extend them to a situation where the system is confined in a quasi-1D quantum wave-guide. We calculate the energy, the effective mass, and the spectral weight of the impurity using a non-perturbative variational approach. We then compare our results to the prediction for the Yang-Gaudin 1D Hamiltonian,   where the impurity interacts with the background Fermi sea through a zero-range Dirac potential 
\be
V_{1D}(z)=g_{1D}\delta (z),
\ee
and for which exact solutions were found using Bethe's Ansatz
\cite{mcguire1966interacting,guan2013fermi} and we clarify under which condition this Hamiltonian provides a good low-dimensional description of the quasi-1D system.

We consider a single impurity immersed in a sea of spin-polarized fermions. We assume that all particles are confined in a cylindrical harmonic potential and we note $\omega_\perp$ the trapping frequency in the $(x,y)$ plane. The particles are free to move in the $z$ direction. We consider a tightly confined regime where, in the absence of interactions, both temperature and  Fermi energy of the Fermi sea are lower than $\hbar\omega_\perp$, which implies that in the absence of coupling, particles populate only the lowest-energy state of the transverse potential. 

We work in the ultracold regime where same-spin fermions do not interact. This also means that the only relevant interaction between impurity-majority atoms can be described using a the zero-range potential $\widehat V_{2b}$ given by 
\be\langle\bm k'|\widehat V_{2b}| \bm k\rangle = \frac{g_0}{L^3} 
\label{interaction}
\ee
 
\noindent where $g_0$ is a coupling constant, $L^3$ is a quantization volume and $\bm k$ and $\bm k'$ represent the relative momenta of the two particles. It is well known that this zero-range Hamiltonian is singular, and the coupling constant must be renormalized in order to obtain finite results. Here, we fix the coupling value by anchoring it to the 1D scattering amplitude. Let's denote $|k_{z,r},n_r,m_{z,r}\rangle$ the eigenstates of the relative motion, where $k_{z,r}$ is the momentum along $z$, $\hbar m_{z,r}$ is the angular momentum along that direction and $\hbar\omega_\perp(n_r+1)$ is the energy in the transverse plane. In this basis, the matrix elements of the two-body interaction satisfy the equation

\be
\langle k_z,n_r,m_{z,r}|\widehat V_{\mathrm 2b}|k'_z,n'_r,m'_{z,r}\rangle=\frac{g_0}{2\pi\ell_\perp^2 L}\delta_{m_{z,r},0}\delta_{m'_{z,r},0},
\ee
where $\ell_\perp=\sqrt{\hbar/m\omega_\perp}$ is the typical size of the ground state of the transverse motion. 
 After a straightforward calculation \cite{SuppMat}, the 1D-T-matrix for particles in the ground state of the transverse harmonic confinement can be written as
\be
\begin{split}
\frac{1}{T(E)}=&L\Bigg[\frac{2\pi\ell_\perp^2}{g_0}-\frac{1}{L}\sum_{p,n>0}\frac{1}{p^2/m+2\hbar\omega_\perp n}+\\
&\frac{1}{2^{3/2}\hbar\omega_\perp\ell_\perp}\left[\zeta(1/2,-E/2\hbar\omega_\perp)-\zeta(1/2)\right]
\Bigg],
\end{split}
\ee
where $E$ is the total kinetic energy of the incoming particles and $\zeta(\alpha,s)$ and $\zeta(\alpha)$ are respectively  Hurwitz's and Riemann's zeta functions. This expression is formally similar to that found in \cite{olshanii1998atomic} if we identify the constant term with the coupling constant $g_{1D}=-2\hbar^2/ma_{1D}$ calculated in this earlier work. More precisely, the 1D-scattering length is related to the 3D scattering length $a_{3D}$ by 

\be
a_{1D}=-\frac{\ell_\perp^2}{a_{3D}}\left[1+\frac{\zeta(1/2)}{\sqrt{2}}\frac{a_{3D}}{\ell_\perp}\right].
\label{Eq:1DScatteringLength}
\ee
Below, we will thus use the prescription

\be
\frac{1}{g_{1D}}=\frac{2\pi\ell_\perp^2}{g_0}+\frac{1}{L}\sum_{p,n>0}\frac{1}{p^2/m+2n\hbar\omega_\perp}.
\ee
in order to cure the singularity appearing in the two terms in the rhs of this equation in the zero-range limit \footnote{Note that the value of the 1D-scattering length can be found explicitly using e.g. a separable Gaussian interacting potential.}.

\begin{figure}
\includegraphics[width=\columnwidth]{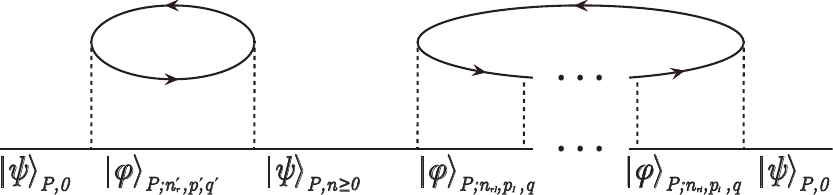}
\caption{Diagrammatic representation of a typical process appearing in the calculation of the polaron resolvent operator. The polaron starts in the initial state $|\psi\rangle_{P,0}$. The first interaction creates a particle-hole pair $|\varphi\rangle_{P;n^\prime_r,p^\prime,q^\prime}$. The second interaction (in this particular case) annihilates the pair, while the polaron can be promoted in an arbitrary state  $n \geq 0$ of the transverse harmonic confinement. Since the Fermi sea is back to its intial state and the particle momentum is $P$, we call this state a "forward scattering" state. The next interaction creates a new hole of momentum $q$, while during the polaron and the particle exchange momentum and can be transferred towards excited states during the following interactions. The last interactions brings the system back to its initial state.}
\label{Fig:diagrams}
\end{figure}

Let's denote $\widehat a_{k_z,n,m_z}$ ($\widehat b_{k_z,n,m_z}$) the annihilation operator of a majority (minority) atom in a single-particle state $|k_z,n,m_z\rangle$. Assuming that the impurity has an initial momentum $P$ along $z$, we calculate its energy by looking for the poles of the matrix element of the resolvent operator $\widehat G(z)=1/(z-\widehat H)$ on the state $|\psi\rangle_{P,0}=b^\dagger_{P,0,0}|{\rm FS}\rangle$ describing an impurity of momentum $P$ in the presence of a non-interacting Fermi Sea of majority atoms $|{\rm FS}\rangle=\prod_{|k_z|<k_F}\widehat a^\dagger_{k_z,0,0}|0\rangle$, where $|0\rangle$ is the vacuum. We calculate it using a partial resummation of Lippman-Schwinger's expansion 
\be
\widehat G=\widehat G_0+\widehat G_0\widehat V \widehat G_0+...
\label{Greenen}
\ee
where $\widehat G_0$ is the resolvent operator of the non-interacting system and $\widehat V$ is the interaction between the impurity and the Fermi sea. 

Following \cite{chevy2006upa}, we approximate Eq.~(\ref{Greenen}) to include at most one particle-hole excitation of the Fermi sea. This implies two different types of propagation for the polaron. The first type is a ``forward-scattering process"  where the impurity keeps its axial momentum, does not create any excitation of the Fermi sea, but is virtually promoted to an excited state of the transverse potential. We write the associated state as $|\psi\rangle_{P,n}=\widehat b^\dagger_{P,n,0}|FS\rangle$. The second class of states corresponds to the formation of a particle-hole pair excitation after interaction with the impurity. We assume that this interaction does not affect the center of mass state of the interacting pair. This enforces a partial Kohn theorem to the impurity/particle pair \cite{Kohn1965New} and the corresponding states can be written as $|\varphi\rangle_{P;n_r,p,q}\propto\widehat C^\dagger_{P+q,0,0;p,n_r,0}\widehat a_{q,0,0}|FS\rangle$, where we introduced the operator $\widehat C_{K_z,N,M_z;k_{z,r},{n_r},m_{z_r}}$ corresponding to the annihilation of a majority/minority pair of atoms, whose center of mass (resp. relative motion) is in state $|K_z,N,M_z\rangle$ (resp. $|k_{z,r},n_r,m_{z,r}\rangle$). 

We denote $G'_0$ and $B$ the respective contributions of these two classes of propagation  (See Appendix~\ref{sec:di} for explicit expressions and \cite{SuppMat} for their derivation).
In Fig.~\ref{Fig:diagrams} we show a specific diagram which includes both types of propagation.


 A key point of the calculation is that in the zero range limit, $g_0$ becomes vanishingly small. It means that the only diagrams contributing to the final result are associated with divergent sums over momenta. This criterion, for instance, excludes the interactions of the impurity with holes or successive forward scattering where the impurity interacts several times with the Fermi sea without creating any hole.     

The resummation of the series yields the following expression

\begin{eqnarray}
\null_{P,0}\langle\psi|\widehat G(z)|\psi\rangle_{P,0}&=&G_0+G_0^2\sum_{n=0}^\infty (G_0'B)^n\\
&=&\frac{1+ B \left( G_0 - G^{\prime}_0 \right)}{\frac{1}{G_0} - B \frac{{G}^{\prime}_0}{G_0}}.
\label{Green}
\end{eqnarray}
Interestingly, the self-energy energy of the polaron can be written in this approximation as
\be
\Sigma= \frac{B}{1+(G_0-G^{\prime}_0)B},
\ee
which differs from the more common expression $\Sigma=B$ obtained for instance for 3D systems \cite{combescot2007nsh}. The origin of the difference is the contribution of virtual excitations in $G_0'$. Without them, we would have $G_0'=G_0$, and the self-energy would be given solely by the particle-hole bubble. 

By definition, the poles of the resolvent operator correspond to the eigenstates of the Hamiltonian. The energy $z=E(P)$ of a polaron of momentum $P$ is therefore solution of the equation
\be
G'_0(z)B(z)=1.
\ee
At low momenta, we expand the energy as
\be
E(P)=E_0+\frac{P^2}{2m^*}+...
\ee
\noindent where $E_0$ is the energy of a polaron at rest, and $m^*$ is its effective mass. 

\begin{figure}
\includegraphics[width=\columnwidth]{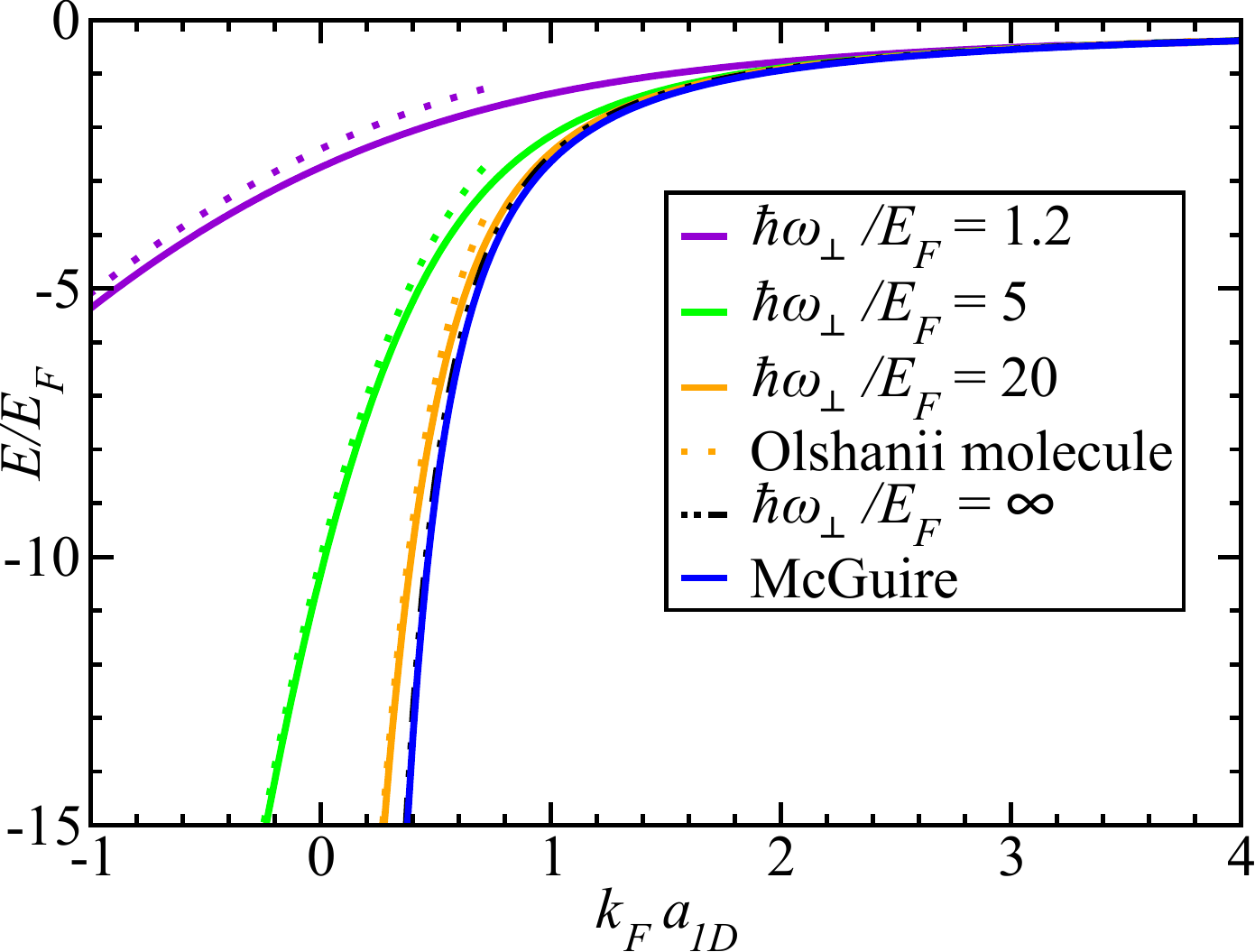}
\includegraphics[width=\columnwidth]{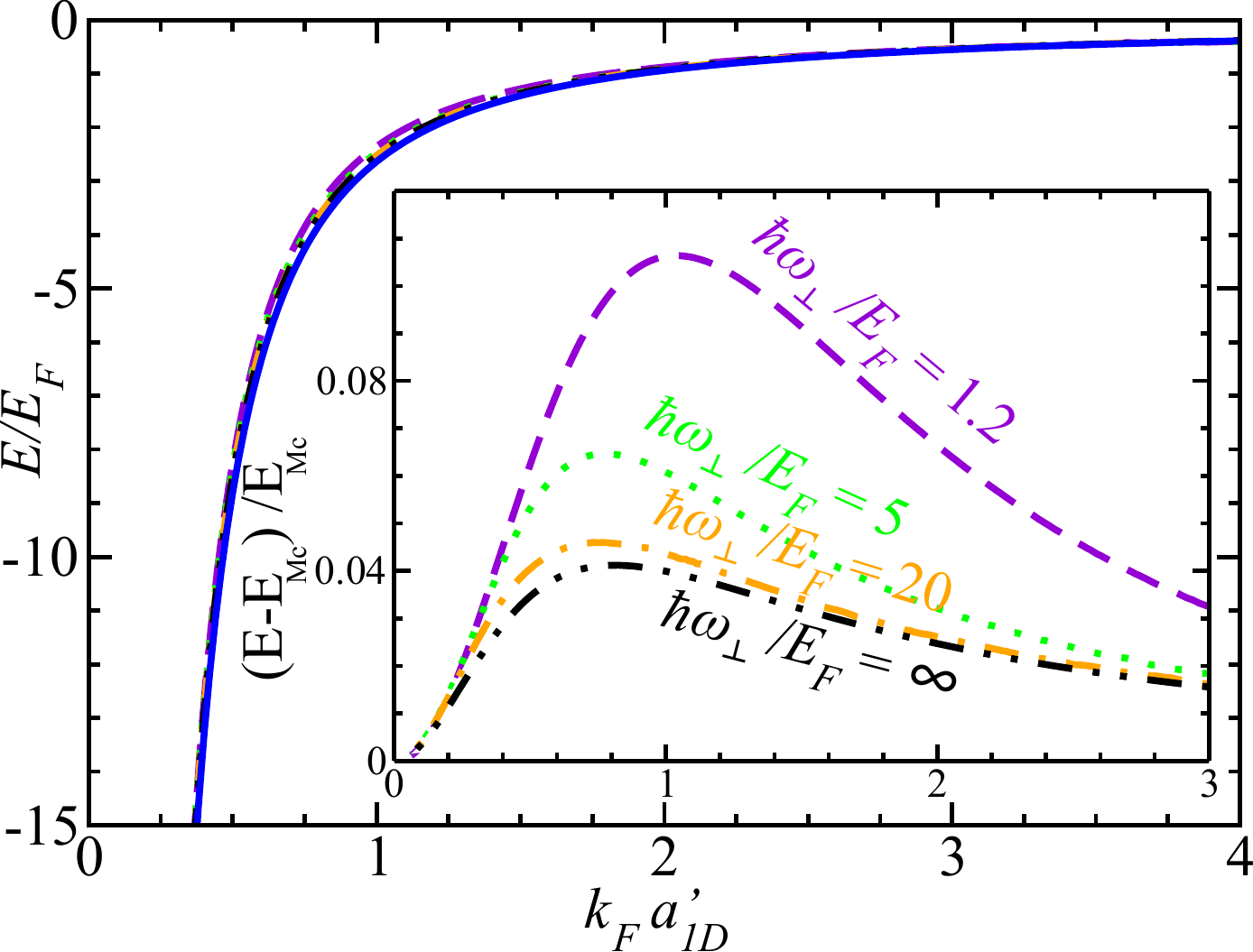}
\caption{Upper panel. Polaron energy vs $k_F a_{1D}$ for several values of the transverse trapping potential. We observe a non-universal behavior that does not coincide with the truly 1D prediction. Lower panel: same, but represented as a function of $k_F a'_{1D}$ where the effective scattering length $a'_{1D}$ is always positive and chosen to recover the energy of the two-body bound state (see text). In this case all curves collapse on McGuire's result, Eq. (\ref{Eq:McGuire}) (blue curve). The inset displays  the relative difference between the quasi- and purely-1D predictions. In both panels the transverse trapping frequencies correspond respectively to $\hbar\omega_\perp/E_F=1.2$ (purple dashed); 5 (green dotted); 20 (orange dash-dotted) and $\infty$ (black dashed double-dotted).}
\label{Fig:Energy}
\end{figure}

We first discuss the dependence of $E_0$ on the different physical parameters of the problem and we compare it to the prediction for a true 1D system (Fig.  \ref{Fig:Energy}). For the sake of simplicity we consider that the impurity and the majority atoms correspond to two different spin states of the same fermionic species. Consequently, they share the same mass and  are confined by the same transverse harmonic potential of frequency $\omega_\perp$. For a true 1D system characterized by a 1D scattering length $a_{1D}$, the energy of the polaron is given by the following expression \cite{mcguire1966interacting}
\be
\frac{E}{E_F}=-\frac{2}{\pi}\left[u-\frac{\pi}{2}u^2+(1+u^2)\arctan(u)\right],
\label{Eq:McGuire}
\ee
with $u=-1/(k_Fa_{1D})$. In order to compare between the quasi- and the truly 1D situations, one needs a prescription fixing the value of $a_{1D}$ using the real life system physical parameters ($a_{3D}$, transverse confinement etc...). The scattering of two distinguishable particles in a harmonic quantum wire was studied in \cite{olshanii1998atomic} and the corresponding 1D scattering length is given by Eq. \ref{Eq:1DScatteringLength}.  The dependence of $E/E_F$ with $k_F a_{1D}$ for several values of the parameter $\hbar\omega_\perp/E_F$ is displayed in Fig. (\ref{Fig:Energy}) and we compare it to McGuire's prediction for a purely 1D system. We observe first that both quasi- and truly 1D systems agree in the weakly attractive limit $k_F a_{1D}\rightarrow\infty$. Strictly speaking, the agreement between these two regimes is valid only at the mean-field level. Indeed at second order in interactions, virtual transitions towards excited states of the transverse potential give rise to finite range corrections and emergent three-body interactions that can be calculated directly \cite{Chevy2022AchievingFermions} and are not included in (\ref{Eq:McGuire}).

When interactions are increased, we observe that the quasi- and purely 1D predictions diverge rapidly. One striking feature is the existence of the attractive polaron only for positive $a_{1D}$ in the purely 1D regime, while this branch exists for all values the 1D scattering length in the quasi-1D case. This discrepancy comes from the fact that in the strongly attractive limit, the impurity binds with an atom from the Fermi sea to form a molecule. While in a quasi-1D system, the bound state exists for arbitrary values of $a_{1D}$ \cite{olshanii1998atomic,moritz2005confinement}, its stability domain is restricted to positive values of the 1D scattering length in the Yang-Gaudin Hamiltonian.  

 This discrepancy can be partially resolved by modifying the prescription for the definition of the 1D scattering length. Indeed, as proposed in previous works in quasi-1D or 2D \cite{liao2009spin,boettcher2016equation}, one can define a new 1D scattering length -- that we note here $a'_{1D}$ -- by fixing the binding energy of the 1D two-body problem to the value predicted in the quasi-1D geometry. More precisely, if we note $E_{B}$ the energy of the dimer calculated in \cite{olshanii1998atomic}, we define $a'_{1D}$ by $E_B=-\hbar^2/m{a'_{1D}}^2$. By construction, $a'_{1D}$ is always positive and the variations of the polaron energy as a function of the new parameter $k_F a'_{1D}$ is displayed in the lower panel of Fig \ref{Fig:Energy}. In the weakly interacting limits, $a_{1D}$
and $a'_{1D}$ coincide and the agreement with McGuire's 1D prediction is thus maintained. In the strongly interacting limit, the 1D-polaron turns into a dimer. The purely and quasi-1D regimes thus yield by construction the same result, $E_0\simeq -\hbar^2/m {a_{1D}'}^2$. In between, we obtain a nearly perfect collapse of the different curves on that of the purely 1D prediction. 

\begin{figure}
\includegraphics[width=\columnwidth]{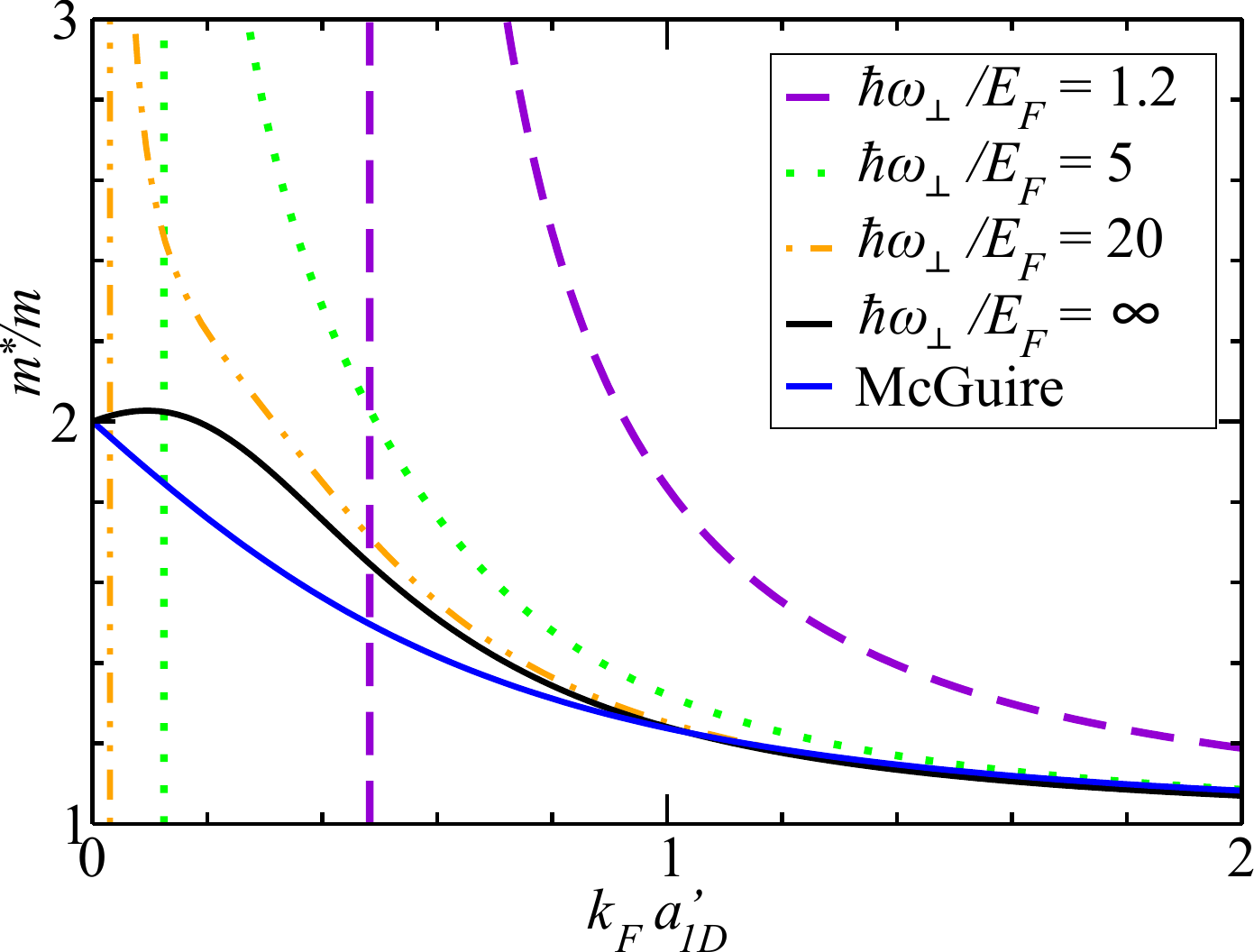}
\caption{In the purely 1D system, the effective mass $m^*$ varies from $m$ to $2m$. In quasi 1D, we observe a divergence of the effective mass in the strongly attractive limit. This divergence is the indirect signature of a breakdown of the polaronic ansatz that is no longer a local energy minimum. This suggests the existence of a discontinuous transition to a molecular state.}
\label{Fig:EffectiveMass}
\end{figure}

This quasi-universal behavior does not apply to all physical quantities beyond the energy of the ground state. A first counter-example was already pointed out in \cite{mora2004atom,mora2005three,Mora2005Four-bodyGas} where it is shown that the atom-dimer and dimer-dimer scattering lengths changes sign when varying the 3D scattering length, contrary to the purely-1D prediction. Here, we illustrate this breakdown in the case of the effective mass $m^*$ of the quasi-particle. Its dependence on interaction (characterized by $k_F a'_{1D}$ and confinement strengths) is displayed in Fig. (\ref{Fig:EffectiveMass}). Contrary to the energy, the parametrization of the scattering length using $a'_{1D}$ does not yield a universal curve. Indeed, while for purely 1D systems, the exact solution obtained using Bethe ansatz predicts a smooth crossover between the mass $m$ of a free atom to that of a dimer ($2m$) \cite{mcguire1966interacting,guanPRA2016}, we see that the polaronic variational state predicts a divergence of the effective mass in the strongly attractive limit. This behavior is reminiscent of the 3D-polaron where a similar divergence occurs and was interpreted as a side effect of the polaron-molecule transition occurring in the strongly attractive limit $k_F a_{3D}\rightarrow 0^+$.

\begin{figure}
\includegraphics[width=\columnwidth]{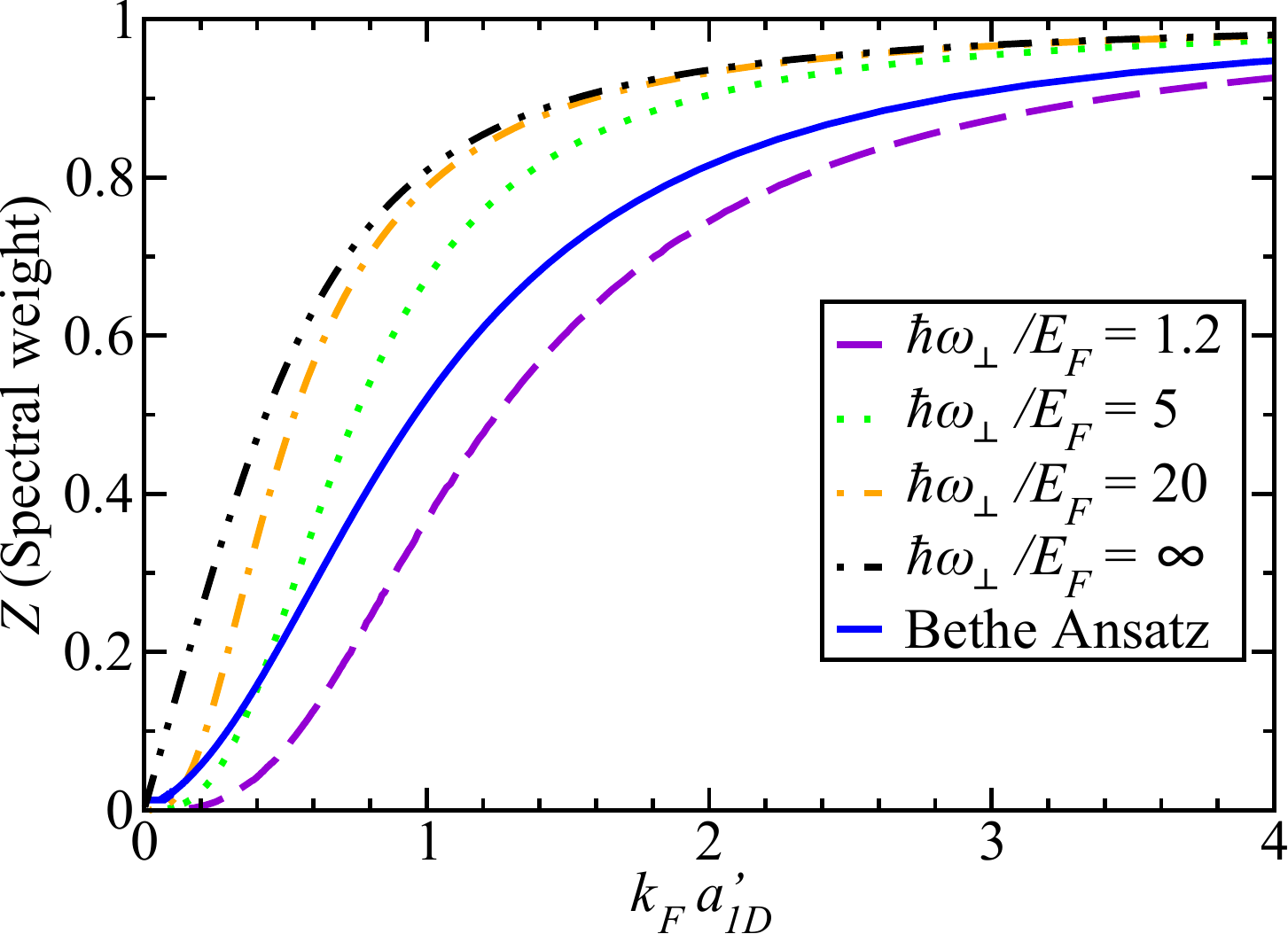}
\caption{Spectral weight of the quasi-1D polaron for a finite size system containing $N=100$ atoms.}
\label{Fig:Residue}
\end{figure}

Finally, we discuss the evolution of the spectral weight $Z$ of the impurity. It corresponds to  the residue of the resolvent operator at the polaron energy and is given by 

\be
Z = \frac{B(E_0)}{E_0\left[1-\frac{\partial}{\partial z} \left(B \frac{G^{\prime}_0}{G_0}\right)_{z=E_0}\right]}
. 
\label{Eq:Z}
\ee

Experimentally, this quantity is accessible through the measurement of the Rabi oscillation frequency between two spin states of the impurity \cite{kohstall2011metastability}. In the Yang-Gaudin model, the spectral weight vanishes in the thermodynamic limit, due to Anderson's orthogonality catastrophe  \cite{Anderson:PRL1967,castella1993exact,levinsen2015strong, Orso2025Quasi-particlePolaron}. This result contradicts Eq. (\ref{Eq:Z}) as well as the variational study of purely 1D systems reported in \cite{Giraud2009HighlyCase} which both predict a finite value in the thermodynamic limit. However, in real experiments the system contains only a finite number $N$ of  fermions. This means that in practice, the purely Yang-Gaudin model also predicts a finite value for the spectral weight of the impurity. 

In Fig. \ref{Fig:Residue} we display this variational prediction for $Z$ for $N=100$ for different values
of the confinement strength, together with the
purely 1D prediction that we obtained from Bethe-ansatz (see \cite{SuppMat}).
We see that contrary to the thermodynamic limit, the quasi and purely 1D results predict similar finite values for the polaron spectral weight for typical experimental conditions \cite{de2021situ,liao2009spin}.

In conclusion, we have shown that the non-perturbative variational approach presented in this Letter can capture physical phenomena beyond the simple Yang-Gaudin description of quasi-1D systems. Using it, we could describe the effect of virtual excitations towards transverse excited states while keeping under control the UV singularities associated with 3D zero-range interactions. Fig. \ref{Fig:Energy} provides an estimate of the accuracy of a quasi-1D system as a simulator of the Yang-Gaudin model. For a moderately tight confinement $\hbar\omega_\perp/E_F=5$, corresponding to the parameters of the experimental results reported in \cite{de2021situ}, we estimate the error on the energy to be lower than 7\%. However, even if the values of the energy are very similar, the two models can disagree for other features, like the value of the dimer-dimer scatering length or the  existence of a transition to a molecular state whose existence is suggested by the divergence of the effective mass but is absent in the 1D Yang-Gaudin model.

Our calculation is restricted to a single particle-hole pair excitation. This approximation is known to be accurate in arbitrary dimensions for the energy of the ground state but fails to capture other features, such as Anderson's orthogonality catastrophe for a purely-1D mobile impurity \cite{Anderson:PRL1967}. One open question that remains is thus the value of $Z$ at the thermodynamic limit in quasi-1D. Solving this question requires considering multiple particle-hole excitations and could be addressed using diagrammatic Monte-Carlo approaches.

\acknowledgements

The authors acknowledge support from Institut Universitaire de France, PEPR Project Dyn-1D (ANR-23-PETQ-001), ANR Collaborative project ANR-25-CE47-6400 LowDCertif and ANR International Project ANR-24-CE97-0007 QSOFT. 

\bibliography{references.bib}

\newpage
\appendix
\section{Expression of diagrams}
\label{sec:di}

Here we will give the expression for the two classes of diagrams. The $G^{\prime}_0(z)$ is the summation of all green propagators where the majority atoms are in the Fermi sea $|FS\rangle$, and the polaron of momentum $P$ is in any state of the harmonic oscillator. If we write this summation in the center of mass and evaluate the matrix elements of the interaction in the frame of the center of mass: 

\begin{eqnarray}
G^{\prime}_0(z)= \sum_{n_r \geq 0} \frac{1}{4^{n_r}} \frac{1}{z - P^2/2m- 2 n_r \hbar \omega_{\perp}},
\label{forwardsimple}  
\end{eqnarray}

\noindent where the sum is over $2n_r$ is the sum over all the states of center of mass with the quantum number $m_z=0$. 

The bubble diagram $B(z)$ represents all the possible diagrams when a particle from the hole of momentum $q$ is created. As the hole can have a momentum $q \in [-k_F,k_F]$ the bubble propagator can be recast as a sum of $q$:    

\begin{eqnarray}
B(z) &=&  \sum_{q < k_F} \sum^{\infty}_{k=1} \left(\frac{g_0}{2\pi\ell_\perp^2 L}\right)^{k+1} b^k(z,q) \nonumber \\ &=&  \sum_{|q|<k_F} \frac{ 1 }{\frac{2\pi\ell_\perp^2 L}{g_0} - b(z,q)}, 
\label{bubbleee}
\end{eqnarray}

\noindent where $b(z,q)$ represents the propagation of all the possible particle-hole excitations where the hole has a momentum $q$, the particle has a momentum $p$, while the momentum conservation law implies that the polaron has a momentum $P+q-p$. After renormalizing the coupling constant $g_0$ one gets after some straightforward algebra \cite{SuppMat}

\begin{widetext}
\begin{align}
\frac{2\pi\ell_\perp^2}{g_0} - \frac{b\left( z , q \right)}{L}=&\frac{1}{g_{1D}} +  \frac{1}{L}\sum_{n_r>0, p} \left[\frac{1}{\varepsilon(p,q;P)+ \hbar \omega_{\perp} 2 n_r-z}-\frac{1}{p^2/m+\hbar\omega_\perp 2 n_r} \right] \nonumber \\
&+\frac{1}{L}\sum_{n'_2>0,|p|<p_F}\frac{1}{4^{n'_2}}\frac{1}{z-\varepsilon(p,q;P)-2\hbar\omega_\perp n'_2} \nonumber \\ 
&+\frac{1}{L}\sum_{|p|>p_F}\frac{1}{\varepsilon(p,q;P)-z},
\label{particleholesimple}  
\end{align}
\end{widetext}
 
\noindent where $\varepsilon(p,q;P) = \left(p^2 - p\left( q+P \right) +Pq+P^2/2 \right)/m$. 


\clearpage
\onecolumngrid

\begin{center}
\huge
Supplemental Material
\normalsize
\end{center}

\setcounter{equation}{0}
\setcounter{figure}{0}
\setcounter{table}{0}
\setcounter{page}{1}
\makeatletter

These supplementary materials provide the technical details underlying the work presented in the article. In the first section, we discuss the renormalization of the two-body interaction in a quasi-1D waveguide, outlining the key steps in its derivation. The second section focuses on the diagrammatic resummation approach used to obtain the final expression for the Green's function, examining the contributions from different diagrammatic components. Finally, in the third section, we describe the calculation of the exact 1D spectral function of a polaron, following the approach introduced in the main text.

\section{The two-body problem in a quantum-wave guide}

This section is devoted to presenting an alternate method to \cite{olshanii1998atomic2} yielding a formal expression of the effective quasi-1D coupling constant easily amenable to renormalization in the zero-range limit. 

In this section, we regularize the two-body interaction using separable Gaussian potential 

\begin{eqnarray}
\widehat V_{\mathrm 2b}=\frac{g_0}{L^3}|\chi_\Lambda\rangle \langle\chi_\Lambda|
\label{addedpart} \; ,
\end{eqnarray}
with 
\be
\langle \bm k|\chi_\Lambda\rangle=\chi_\Lambda(\kappa)=e^{-\kappa^2/2\Lambda^2}
\ee
and where the bare coupling constant $g_0$ must be renormalized during our calculations.

{\em Free space T-matrix}: We first calculate the $T$ matrix in free space. After a straightforward resummation of the perturbation series for the T-matrix, we have

$$
\langle\bm k|T|\bm k'\rangle=\frac{1}{L^3}\frac{\chi_\Lambda(\bm k)^*\chi_\Lambda(\bm k')}{\frac{1}{g_0}-\frac{1}{L^3}\sum_{k_1}\frac{|\chi(k_1)^2|}{z-\hbar^2k_1^2/m}},
$$
with $z=\hbar^2 k^2/m+i0^+$. The sum in the denominator is, strictly speaking, divergent in the zero range limit. To cure this singularity, we add and subtract counter, which allow us to recast the T-matrix as

$$
\langle\bm k|T|\bm k'\rangle=\frac{1}{L^3}\frac{1}{\frac{1}{g_{3D}}-\frac{1}{L^3}\sum_{k_1}\left[\frac{1}{z-\hbar^2k_1^2/m}+\frac{1}{\hbar^2 k_1^2/m}\right]},
$$
where
$$
\frac{1}{g_{3D}}=\frac{1}{g_0}+\frac{1}{L^3}\sum_{k_1}\frac{|\chi_\Lambda(k_1)|^2}{\hbar^2k_1^2/m}=\frac{1}{g_0}+\frac{m\Lambda}{8\sqrt{\pi}\hbar^2}.
$$
is the ``true" physical coupling constant associated with the zero-energy value of the scattering amplitude.

{\em Scattering in a wave-guide}. We now consider the scattering in a wave-guide. Thanks to Kohn's theorem, the motions of the center of mass and relative coordinates are decoupled. Moreover, by rotational symmetry, if the atoms start in the transverse ground state, their relative angular momentum along the symmetry axis of the wave-guide is zero throughout the scattering. The matrix elements of $V$ in the basis $|q,n,m_z\rangle$ of the relative motion are given by
$$
\langle q,n,0|V_{\mathrm 2b}|q'n',0\rangle=\frac{g_0}{2L\ell_\perp^2}\mu_n^*\mu_{n'},
$$
with 
$$
\mu_n=\frac{1}{L}\sum_{\bm\kappa}\langle\bm\kappa|n,0\rangle\chi_\Lambda(\kappa)=\frac{1}{\sqrt{\pi}}\frac{1}{1+1/(2\Lambda^2 \ell_\perp^2)}X(\Lambda \ell_\perp)^n.
$$
Here $n=2 n_1$ is even since we consider only the $m_z=0$ channel, $\ell_\perp=\sqrt{\hbar/m\omega_\perp}$ an $X(u)=(1-1/(2u^2)/(1+1/(2u^2)=1-1/u^2+O(u^{-4})$.

We see that the potential is still factorized in the basis of the eigenstates of the transverse oscillator. We can therefore once again resume the series expansion of the T-matrix and we have

$$
\langle q,n=0, m_z=0|T|q',n'=0,m'_z=0\rangle=\frac{g_0 |\mu_0|^2/\left(2L\ell_\perp^2\right)}{1-\frac{g_0}{2 \ell_\perp^2L}\sum_{q_1,n_1}\frac{|\mu_{n_1}|^2}{z-q_1^2/2-2n_1}}.
$$

Following the same procedure as in free space, we get rid of the convergence factors by introducing counter terms, allowing us to get rid of the cut-off $\Lambda$
\be
\begin{split}\frac{1}{
\langle q,0|T|q',0\rangle}=2\ell_\perp^2 L\Bigg[&\frac{\pi}{g_0}+\frac{1}{2\ell_\perp^2L}\sum_{q_1,n_1>0}\frac{X(\Lambda)^{2n_1}}{\hbar^2q_1^2/m+2\hbar\omega_\perp n_1}\\&-\frac{1}{2\ell_\perp^2 L}\sum_{q_1,n_1>0}\left(\frac{1}{z-\hbar^2 q_1^2/m-2\hbar\omega_\perp n_1}+\frac{1}{\hbar^2 q_1^2/m+2\hbar\omega_\perp n_1}\right)\\&-\frac{1}{2\ell_\perp^2 L}\sum_{q_1}\left(\frac{1}{z-\hbar^2 q_1^2/m}\right)\Bigg].
\end{split}
\label{Eq:T-Matrix}
\ee
(note that we have to distinguish the case $n_1=0$ because for $z=0$ and $n_1=0$, the sum over $q$ is infrared divergent, and we can not use it as a valid counter term).

Taking the zero-range limit $\Lambda\rightarrow\infty$, we recover the expression  found in \cite{olshanii1998atomic2}

\be
\frac{1}{
\langle q,0|T|q',0\rangle}=L\Bigg[\frac{1}{g_{1D}}+\frac{i}{q}+F(q)\Bigg],
\label{Eq:Quasi1DTMatrix}\ee
with $F(q)=\zeta(1/2,1-\hbar q^2/(2m\omega_\perp))-\zeta(1/2)$
where $\zeta(s,z)$ is Hurwitz's zeta function  and where the the 1D-coupling constant $g_{1D}$ is given by $g_{1D}=-2\hbar^2m/a_{1D}$, with 

$$
\frac{a_{1D}}{\ell_\perp}=-\frac{1}{2}\left(\frac{\ell_\perp}{a_{3D}}+\zeta(1/2)\right),
$$

Interestingly, identification with Eq. (\ref{Eq:T-Matrix}) also yields the formal expression 

\be
\frac{1}{g_{1D}}=\frac{2\pi\ell_\perp^2}{g_0}+\frac{1}{L}\sum_{q,n_1>0}\frac{1}{\hbar^2 q_1^2/m+2\hbar\omega_\perp n_1}
\label{Eq:Renormalization}
\ee
that will be used to regularize the quasi-1D many-body problem.

\section{Resummation of the resolvent operator for the quasi-1D polaron problem}

In this work, we were using the well-known formalism  of the resolvent operator defined as:

\begin{eqnarray}
\hat{G}\left( z \right) = \frac{1}{z - \hat{H}}
\label{Introgreen}
\end{eqnarray}

\noindent to determine the energy, the effective mass, and the spectral weight of the free polaron state. More precisely, we were looking at its projection on the state $|\psi\rangle_{P,0}=b^\dagger_{P,n,0}|FS\rangle$ describing an impurity of momentum $P$ in the presence of a non-interacting Fermi Sea of majority atoms $|FS\rangle$:

\begin{eqnarray}
G\left( z \right) = _{P,0}\langle\psi| \hat{G} \left( z \right) |\psi\rangle_{P,0}
\label{addedEq0071}
\end{eqnarray}

\noindent The Hamiltonian $\hat{H}$ from equation (\ref{Introgreen}) is the sum of the free Hamiltonian  $\hat{H}_0$ describing an ensemble of non-interacting particles confined in a harmonic cylindrical potential and of the interaction term $\hat{V}$ which we are going to treat variationally using a partial resummation of Lippman-Schwinger's expansion.

\begin{figure*}
\centering
\includegraphics[width=0.8\textwidth]{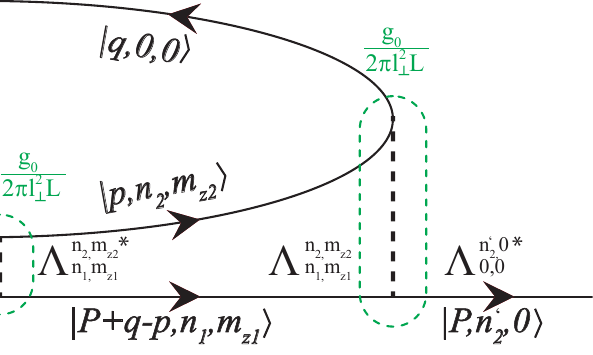}
\caption{Diagrammatic representation of the different contribution around an interaction vertex highlighted in green. On the left of the vertex the polaron propagates in an particle-hole excited state while on the left the polaron is in a forward scattering state. This figure also shows how the projector on the relative frame intervene around the vertex.}
\label{Fig:diagrams}
\end{figure*}

In the variational approach followed here, each term of the expansion is represented by a diagram similar to the one displayed in Fig. \ref{Fig:diagrams} of the main text, where each line represents a particular in a state $|p,n,m_z\rangle$ of the trap. 

If consider two states $|\psi\rangle$
and $|\psi'\rangle$ of the relative motion of the two-body problem, the matrix element of the two-body potential Dirac is simply 

\be
\langle \psi|\widehat V_{\mathrm 2b}|\psi'\rangle=g_0\psi(\bm r=0)^*\psi'(\bm r=0).
\ee

If we consider eigenstates of the cylindrical harmonic potentials $|p_r,n_r,m_{z,r}\rangle$, then this means that the potentials couple only state with 0 angular momentum along the $z$ axis, and we have in this case

\be
\langle p,n_r,m_{z,r}|\widehat V_{\mathrm 2b}|p',n'_r,m'_{z,r}\rangle=\frac{g_0}{2\pi\ell_\perp^2 L}\delta_{m_{z,r},0}\delta_{m'_{z,r},0}.
\ee

In a second quantization form, this operator can be written as 

\be
\widehat V_{\mathrm 2b}=\frac{g_0}{2\pi\ell_\perp^2L}\sum_{\substack{P,N,M_z\\p,n_r\\p',n'_r}}\widehat C^\dagger_{P,N,M_z;p,n_r,m_{z,r}=0}\widehat C_{P,N,M_z;p',n'_r,m_{z,r}=0},
\ee
where $\widehat C_{P,N,M_z;p,n_r,m_{z,r}}$ annihilates a pair of particles in the state $|P,N,M_z;p,n_r,m_{z,r}\rangle$ of the center of mass/relative motion basis. 

We assume that the pairs created in the polaron dynamics are contained in the ground state of the transverse confinement. We therefore restrict the sum to $N=M_z=0$ and we therefore have 

\be
\widehat V_{\mathrm 2b}\simeq\frac{g_0}{2\pi\ell_\perp^2L}\sum_{\substack{P,p,n_r\\p',n'_r}}\widehat C^\dagger_{P,0,0;p,n_r,m_{z,r}=0}\widehat C_{P,0,0;p',n'_r,m_{z,r}=0},
\ee

Let's now decompose the pair annihilation operators on a decoupled basis. We have then 

\be
\widehat C_{P,N,M_z;p,n_r,m_{z,r}}=\sum_{\substack{p_1,n_1,m_{z,1}\\p_2,n_2,m_{z,2}}}\langle P,N,M_z;p,n_r,m_{z,r}|p_1,n_1,m_{z,1};p_2,n_2,m_{z,2}\rangle \widehat a_{p_1,n_1,m_{z,1}}\widehat b_{p_2,n_2,m_{z,2}}
\ee

This means that in a diagrammatic representation, each vertex is associated with a coupling $g_0/(2\pi \ell_\perp^2L)$ and each propagator of a pair of particles in state $|p_1,n_1,m_{z,1};p_2,n_2,m_{z,2}\rangle$ will be associated with an amplitude 

\be
\Lambda_{n_1,m_{z,1}}^{n_2,m_{z,2}}=\langle P,0,0;p,n_r,0|p_1,n_1,m_{z,1};p_2,n_2,m_{z,2}\rangle
\ee
at each end. Note that  the non-zero matrix elements satisfy the conservation laws $P=p_1+p_2$, $p=(p_1-p_2)/2$, $n_r=n_1+n_2$ and $m_{z,1}+m_{z;2}=0$ and thanks to the properties of plane waves, the $\Lambda$'s then do not depend on the value of the momenta $p_1$ and $p_2$. 

Let's first consider the 'forward scattering' lines where the impurity propagates on top of a full Fermi sea. We thus have necessarily $n_1=m_{z,1}=0$, and by conservation of angular momentum $m_{z,2}=0$. The contribution to the resolvent is thus

\be
    G'_0(z)=\sum_{n_2}\frac{{\Lambda_{0,0}^{n_2,0}}^*\Lambda_{0,0}^{n_2,0}}{z-P^2/2m-\hbar \omega_\perp n_2}.\\
\ee 

Using standard harmonic oscillator algebra, we have 

\begin{eqnarray}
\ket{N=0, M_z=0} \otimes \ket{ n_r=2n_r^{\prime}, m_{z,r}=0} = \frac{1}{2^{n_r^{\prime}}} \sum_{k k^{\prime}} \sqrt{\binom{n_r^{\prime}}{k} \binom{n_r^{\prime}}{k^{\prime}}} \ket{k + k^{\prime}, k - k^{\prime}}_1 \otimes \ket{2 n_r^{\prime} - k - k^{\prime}, k^{\prime} - k}_2.
\label{Eq006}
\end{eqnarray}
 In the present situation, we have $n_1 = 0$, $m_{z,1} = 0$  and $m_{z,2} = 0$. The fact that the angular momentum is zero means that the quantum number $n_2$ is even and we thus take $n_{2} = 2n_2^{\prime}$, which leads to the identity

\begin{eqnarray}
\Lambda_{0,0}^{n_2=2n'_2,0} = \frac{1}{2^{n_2^{\prime}}},
\label{Eq007}
\end{eqnarray}
hence 

\be
    G'_0(z)=\sum_{n_2=2n'_2}\frac{1}{4^{n'_2}}\frac{1}{z-P^2/2m-2\hbar \omega_\perp n'_2}\\
\ee

We now turn to the calculation of the contribution to $G$ of the bubble diagrams.

Let's note first $b(z,q)$ the contribution of the particle-hole excitations to the propagator. 

Using this definition, the contribution of a full particle-hole bubble will simply be given by the geometric series
    
    \begin{equation}
B(z)=  \sum_{q < k_F} \sum^{\infty}_{k=1} \left(\frac{g_0}{2\pi\ell_\perp^2 L}\right)^{k+1} b^k(z,q) =  \sum_{|q|<k_F} \frac{ 1 }{\frac{2\pi\ell_\perp^2 L}{g_0} - b(z,q)}, 
\label{bubbleee}
\end{equation}
where we have used the fact that in the zero range limit, we have $g_0 b/2\pi\ell_\perp^2 L\rightarrow 1$ in order to compensate for the renormalization of the bare coupling constant $g_0$.

$b$ can be expressed as:

\begin{eqnarray}
 b(z,q)  =  \sum_{\substack{(n_1,m_{z,1},p)\in \mathcal{D}\\ (n_2,m_{z,2})} } \frac{\left|\Lambda_{n_1,m_{z,1}}^{n_2,m_{2}}\right|^2 }{z -\varepsilon(p,q;P) - \hbar \omega_{\perp} \left(n_1 +n_2 \right)}
\label{particleholesimple}  
\end{eqnarray}
Here $(n_1,m_{z,1})$ and $(n_2,m_{z,2})$ are the quantum numbers of the fermion and the impurity respectively. The sum over $(n_2,m_{z,2})$ is unrestricted because the polaron is not affected by Pauli principle.   We introduce $\mathcal{D}$ to denote all the possible values of $n_1$ and $p$, accounting for Pauli Principle, since for $n_1=0$, the momentum of the fermions is restricted to $|p| > k_F$. Finally, $\varepsilon(p,q;P)=p^2/2m+(q+P-p)^2/2m-q^2/2m$ is the kinetic energy of the polaron and the particle-hole pair.

To calculate $b$ we would like to interpret the sum over $(n_1,m_{z,1}; n_2,m_{z,2})$ as a completeness relation that would allow us to turn this expression into a sum over the center of mass/relative degrees of freedom but Pauli Principle prevents such manipulation of this expression. We can nevertheless fulfill this objective by decomposing the sum in the following way:

        \begin{equation}
\sum_{\substack{(n_1,m_{z,1},p)\in \mathcal{D}\\ (n_2,m_{z,2})} }  =  \sum_{\substack{(n_1,m_{z,1}) \\(n_2,m_{z,2})\\ p }}   - \sum_{\substack{n_1=m_{z,1}=0\\ (n_2,m_{z,2})\\|p| < p_F}}
\label{summationseparation}  
\end{equation}
The first term can be recast as a sum over the relative and center of mass degrees of freedom, while the second one involves only fermions in the ground state Fermi sea and is thus similar to the forward-scattering contribution.         

After some straightforward algebra, we thus have

\begin{equation}
\frac{2\pi\ell_\perp^2}{g_0} - \frac{b\left( z , q \right)}{L}=\frac{2\pi\ell_\perp^2}{g_0} +  \frac{1}{L}\sum_{n_r, p} \frac{1 }{\varepsilon(p,q;P)+ \hbar \omega_{\perp} 2 n_r-z}+\frac{1}{L}\sum_{n'_2,|p|<p_F}\frac{1}{4^{n'_2}}\frac{1}{z-\varepsilon(p,q;P)-2\hbar\omega_\perp n'_2}  
\end{equation}

The second term of this expression is divergent. However, this singularity can be cured thanks to the renormalization of $g_0$ using Eq. (\ref{Eq:Renormalization}). Expressing $g_0$ using the 1D coupling constant we have indeed

 \begin{equation}
 \begin{split}
\frac{2\pi\ell_\perp^2}{g_0} - \frac{b\left( z , q \right)}{L}=&\frac{1}{g_{1D}} +  \frac{1}{L}\sum_{n_r>0, p} \left[\frac{1}{\varepsilon(p,q;P)+ \hbar \omega_{\perp} 2 n_r-z}-\frac{1}{p^2/m+\hbar\omega_\perp 2 n_r} \right]\\
&+\frac{1}{L}\sum_{n'_2>0,|p|<p_F}\frac{1}{4^{n'_2}}\frac{1}{z-\varepsilon(p,q;P)-2\hbar\omega_\perp n'_2}\\ 
&+\frac{1}{L}\sum_{|p|>p_F}\frac{1}{\varepsilon(p,q;P)-z}
\end{split}
\end{equation}


\section{Variational nature of the calculation}

We show here that the energy found in the article provides an upper bound of the energy of the true ground state of the quasi-1D Fermi polaron. 

In our calculation, we split the two body-interaction as 

\be\widehat V_{\mathrm 2b}=\widehat V_{\mathrm 2b}^{(0)}+\widehat V_{\mathrm 2b}^{(1)},\ee
where
\begin{eqnarray}
\widehat V_{\mathrm 2b}^{(0)}&=&\frac{g_0}{2\pi\ell_\perp^2L}\sum_{\substack{P,p,p'\\
n_r,n_r'}}\widehat C_{P,0,0;p',n'_r,0}^\dagger \widehat C_{P,0,0;p,n_r,0}\\
\widehat V_{\mathrm 2b}^{(1)}&=&\frac{g_0}{2\pi\ell_\perp^2L}\sum_{\substack{P,p,p'\\
n_r,n_r'\\(N,M_z)\not = (0,0)}}\widehat C_{P,N,M_z;p',n'_r,0}^\dagger \widehat C_{P,N,M_z;p,n_r,0}
\end{eqnarray}

Let's note $\widehat H_0$ the single body contribution of the Hamiltonian (kinetic energy and trapping potential) and $\widehat H=\widehat H_0+\widehat V_{\mathrm 2b}$ the full quasi-1D many-body Hamiltonian. We note $E_0$ its ground state energy. 

In our calculation we consider the approximate Hamiltonian $\widehat H'=\widehat H+\widehat V_{\mathrm 2b}^{(0)}$. Let's note $|\psi'_0\rangle$ and $E'_0$ its ground state and the associated energy.

Since $E_0$ is by definition the lowest energy of $\widehat H$, we have the inequality 

\be
E_0\le \langle\psi'_0|\widehat H|\psi'_0\rangle=E'_0+\langle\psi'_0|\widehat V_{\mathrm 2b}^{(1)}|\psi'_0\rangle.
\ee
We further note that the mean-value of $\widehat V_{\mathrm 2b}^{(1)}$ can be recast as

\be
\langle\psi'_0|\widehat V_{\mathrm 2b}^{(1)}|\psi'_0\rangle = \frac{g_0}{2\pi\ell_\perp^2 L}\sum_{P,(N,M_z)\not = (0,0}\left\|\sum_{p,n_r}\widehat C_{P,N,M_z;p,n_r,0}|\psi'_0\rangle\right\|^2
\ee
Since in the zero-range limit, $g_0\rightarrow 0^-$, we readily see that $\langle\psi'_0|\widehat V_{\mathrm 2b}^{(1)}|\psi'_0\rangle\le 0$, hence $E_0\le E'_0$.

 Finally, the partial resummation of the resolvent operator that we describe in the previous section  amounts to a variational calculation of the ground state energy  of $\widehat H'$ restricted to the single particle-hole sector. The energy $\tilde E'_0$ that we calculate thus satisfy $\tilde E'_0\ge E'_0$, hence $\tilde E'_0\ge E_0$.  
\section{Bethe ansatz calculation of the spectral weight in the 1D limit}

The Bethe ansatz equations for the Yang-Gaudin model describing $N_\uparrow$ spin up fermions 
interacting via a contact potential of strength $g_{1D}$ with one spin down fermion  are given by (assuming periodic boundary conditions)
\begin{eqnarray}
\frac{k_j-\lambda+i c^\prime}{k_j-\lambda-i c^\prime}&=&e^{i k_j L}, \label{BA1}\\
\prod_{j=1}^{N_\uparrow +1} \frac{k_j-\lambda+i c^\prime}{k_j-\lambda-i c^\prime}&=&1,\label{BA2}
\end{eqnarray}
where  $c^\prime=g_{1D}/2$, $\lambda$ is the spin rapidity parameter and
$\{k_j\}$, with $j=1, \dots , N_\uparrow +1$, are the allowed fermion quasi-momenta \cite{mcguire1966interacting2}.
The ground state corresponds to the choice of the quasi-momenta  that minimize the total energy of the system
\begin{equation}
	E=\sum_{j=1}^{N_\uparrow +1} \frac{\hbar^2 k_j^2}{2m}.
\end{equation}

Let $x_{N_\uparrow+1}$ be the position of the impurity and $y_i=x_i-x_{N_\uparrow+1} \in [0,L]$ the relative distance between the i-th fermion and the impurity, with $i=1, \dots , N_\uparrow$. Then the
 many-body wavefunction takes the form: 
 \begin{equation}\label{F}
 	\psi(x_1,\dots,x_{N_\uparrow},x_{N_\uparrow+1})=F(y_1,\dots,y_{N_\uparrow}) e^{i \sum_{j=1}^{N_\uparrow+1} k_j x_{N_\uparrow+1}},
 \end{equation}
where the multidimensional function $F$  is defined as 
\cite{guanPRA20162,guanPRAerratum2}
 \begin{equation}\label{psi2}
 	F(y_1,\dots,y_{N_\uparrow})=\sum_{\ell=1}^{{N_\uparrow}+1} \xi_\ell \textrm{det}(\tilde A^\ell).
 \end{equation}
 In Eq.(\ref{psi2}) we have set
 \begin{equation}
 	\xi_\ell=\begin{cases} 
 		1 & \textrm{if}\; N_\uparrow\; \textrm{even}  \\
 		(-1)^{\ell+1}  &\textrm{if}\; N_\uparrow\; \textrm{odd}, \\
 	\end{cases}
 \end{equation}
 and the entries of the matrices $\tilde A^\ell$ are given by
 \begin{equation}\label{entries2}
 	(\tilde A^\ell)_{is}=[k_{r} -\lambda-i c^\prime]e^{i k_{r} y_i}.
 \end{equation}
with $r=\ell +s$ if $\ell +s \leq N_\uparrow+1$ and $r=\ell +s-N_\uparrow-1$ otherwise. 

From the knowledge of $\psi$ we can calculate the spectral weight $Z$ of the polaron thanks to the expression
\begin{equation}\label{Z}
	Z=\frac{|(\psi_\textrm{NI},\psi)|^2}{(\psi_\textrm{NI},\psi_\textrm{NI}) (\psi,\psi) },
\end{equation}	 
where $\psi_\textrm{NI}$ is the ground state wave-function in the absence of interaction
and 
\begin{equation}\label{scalprod}
(\psi,\psi_1)=\int \prod_{j=1}^{N+1}dx_j  \psi(x_1 \dots, x_N, x_{N+1})^* \psi_1(x_1 \dots, x_N, x_{N+1})
\end{equation}
represents the overlap between two generic  states. To compute $Z$, we note that 
the function $F$ in Eq.(\ref{psi2}) is a weighted sum of Slater determinants. The scalar product between any two Slater determinants   is equal to the determinant of the matrix containing the single-particle (orbital) overlaps. The latter can be evaluated analytically, while the determinant of the matrix is computed numerically.

\end{document}